\documentstyle[graphicx,preprint]{jpsj} 

\def\runtitle{Photogeneration Dynamics of a Soliton Pair in Polyacetylene} 
\def\runauthor{Y. {\sc Hirano} and Y. {\sc Ono} } 

\title {
Photogeneration Dynamics \\ of \\ a Soliton Pair in Polyacetylene
} 

\author {
Yukio {\sc Hirano}\footnote{E-mail : hirano@ph.sci.toho-u.ac.jp}
and Yoshiyuki {\sc Ono}\footnote{E-mail : ono@polaron.ph.sci.toho-u.ac.jp}
} 

\inst { 
Department of Physics, Toho University, Miyama 2-2-1, Funabashi, 
Chiba 274-8510} 

\recdate {15 June 1998} 

\abst {
Dynamical process of the formation of a soliton pair  
from a photogenerated electron-hole pair in polyacetylene is  
studied numerically by adopting the SSH Hamiltonian. A weak local 
disorder is introduced in order to trigger the formation. Starting 
from an initial configuration with an electron at the bottom of the
conduction band and a hole at the top of the valence band, separated by 
the Peierls gap, the time dependent 
Schr${\rm \ddot{o}}$ndinger equation for the electron wave functions and the 
equation of motion for the lattice displacements are solved 
numerically. After several uniform oscillations of the lattice system 
at the early stage, a large distortion corresponding to a pair of  
a soliton and an anti-soliton develops from a point 
which is determined by the location and type of the disorder. 
In some cases, two solitons run in opposite directions, leaving breather 
like oscillations behind, and in other 
cases they form a bound state emitting acoustic lattice 
vibrational modes.
}

\kword{
polyacetylene, charged soliton, neutral soliton, bond-type disorder,
site-type disorder, breather,  
simulation of soliton dynamics, soliton confinement, photogeneration 
of soliton
}

\begin{document}
\sloppy
\maketitle

\section{Introduction}
Polyacetylene is the simplest conjugated polymer. 
There are many physically interesting phenomena 
in the polyacetylene, such as unusual optical and electronic properties which are not 
seen in other organic polymers. 
Those phenomena have been found in the experimental and theoretical 
studies during the last two decades. 
It was predicted~\cite{bib:wsssd} that a photogenerated electron-hole 
pair would decay within a few picoseconds~\cite{Vardeny,Rothberg1} 
into a pair of solitons. 
This theoretical prediction 
has excited a lot of studies on transient spectroscopy of polymers. 
It has been pointed out that there exist the soliton and the polaron as typical 
elementary excitations in the {\it trans}-polyacetylene~\cite{bib:ssh,bib:rice}.
Infrared and Raman activities of polyacetylene have also been studied 
intensively and vibrational modes induced by the existence of solitons have been 
confirmed~\cite{bib:vardeny,bib:ehren,bib:schaf,bib:weidman}. 
The soliton in polyacetylene is a complex nonlinear excitation created by the 
coupling between the electronic and the lattice degrees of freedom.
There are two kinds of soliton, a charged soliton and a neutral 
soliton~\cite{Rothberg1,Rothberg2}.
The charged soliton has a unit charge, and its spin is zero. As for the neutral soliton, 
it has a spin of magnitude $1/2$ but no charge. 
It has also been confirmed with the computer simulations  
by A.R.Bishop and coworkers~\cite{breather1,breather2} that 
under a certain condition there exists 
a localized excitation called a breather which oscillates with a 
certain frequency. \par
So far many efforts have been devoted to numerical studies of 
dynamical behaviors of solitons and related nonlinear excitations in 
polyacetylene~\cite{bib:hito,bib:toy1,bib:I,bib:II,bib:III,bib:IV}. 
Most studies have focused on the energy difference between different 
static configurations of solitons, polarons or bipolarons. 
However the dynamical process of the photoexcitation of solitons has 
not yet been studied intensively. 
Phillpot et al.~\cite{breather2} carried out dynamical simulations on the 
photo-induced formation of nonlinear excitations. In their work, 
the dynamics of the electron system is treated within so-called 
^^ ^^ adiabatic dynamics''; namely the electronic configurations are 
determined for eigenstates of the instantaneous Hamiltonian at each 
time. If the lattice relaxation process is discussed within this 
treatment, the total energy is not conserved as is well-known.~\cite{bib:I} 
In the present work we report the 
results of numerical simulations for the formation process of a 
soliton-antisoliton pair from a photoexcited electron-hole pair in 
the perfectly dimerized background. 
The electronic configurations are determined for the eigenstates of 
the initial Hamiltonian as done in the series of numerical works on the 
soliton dynamics in an electric field.~\cite{bib:I,bib:II,bib:III,bib:IV} 
In this treatment the total energy is conserved. 
Although the importance of the  
mutual Coulomb interaction among electrons is pointed out, we use the 
standard Su-Schrieffer-Heeger (SSH) model without the Coulomb 
interactions. This is partly because of 
simplicity and partly because we want to concentrate ourselves to the 
formation process of a soliton-antisoliton pair. In the presence of 
the Coulomb interaction, the possibility of an exciton formation 
cannot be excluded. The effect of the electron-electron interaction 
will be treated in the future studies.\par
The optical excitation process of solitons can be decomposed into two 
steps. The first step is the creation of an electron-hole pair by 
photoabsorption. The second step is the decay process of this 
electron-hole pair into a soliton pair. The latter step involves the  
lattice relaxation, and therefore is much slower than the former  
which is a purely electronic process. We discuss only the latter in 
this paper. In principle there might be a coherence between the two 
steps~\cite{Block}. We assume, however, that the coherence can be 
neglected because 
of the large difference in the characteristic time scale. The 
radiative recombination process of the electron-hole pair is also 
excluded from the consideration.
\par
The paper is organized as follows. In the next section the model and 
the method of simulation is explained. The results of simulations are 
described in \S 3. The last section is devoted to concluding remarks.
\section{Model and Method of Simulation}
The procedure of simulation is summarized as follows. 
 At first, we solve a static problem where the system is in the ground 
 state with a perfectly
dimerized chain and with electronic levels just half-filled.
Thus, the half of the electronic states are occupied by the up and down 
spin electrons.
As the next step we create artificially an electron-hole pair by 
moving one electron from the top of the valence band to the bottom of 
the conduction band across the Peierls gap, in order to mimic the 
photo-excitation. The details of this excitation process are 
disregarded, since the actual optical excitation process is considered 
to occur within a very short time. We take this excited state with an 
electron-hole pair as the initial state of the dynamical simulation 
and pursue the time dependence of the lattice displacements and the 
electronic wave functions.\par
In this paper we employ the SSH model which is considered to be a 
standard model for {\it trans}-polyacetylene. 
\begin{eqnarray} 
H =  &- &\ \sum\limits_{n,s} [t_{0}-\alpha (u_{n + 1} - u_{n})](c_{n+1,s}^{\dagger} 
\ c_{n,s} + c_{n,s}^{\dagger}c_{n+1,s}) \nonumber \\ 
&+ & \sum\limits_{n} \frac{K}{2}{\left(u_{n + 1} - u_{n}\right)}^ 2  +
\sum\limits_{n} 
\frac{M}{2} {\dot{u}}_{n}^{2} + H^{\prime}, 
\label{eq:2.1}
\end{eqnarray}
where $t_{0}$ is the transfer integral between the nearest neighbor sites in the 
equidistant lattice, $\alpha$ the electron phonon coupling.
The second and third terms of 
eq.~(\ref{eq:2.1}) represent the lattice potential energy mainly due to 
$\sigma$-bonding and the kinetic energy of ${\rm CH}$-units, respectively: $u_n$ 
the displacement of the $n$-th ${\rm CH}$-unit, $K$ the spring constant 
and $M$ the mass of a ${\rm CH}$-unit. 
 The last term describes the part related to disorders and expressed as 
\begin{equation}
H^{\prime} = V_{\rm s} \sum\limits_{i,s} c_{i,s}^{\dagger} c_{i,s}, 
\label{eq:2.3}
\end{equation}
in the case of site-type disorders, and as
\begin{equation}
H^{\prime} =  - V_{\rm b} \sum\limits_{i,s} (c_{i,s}^{\dagger} c_{i+1,s} 
+ c_{i+1,s}^{\dagger}c_{i,s}),  
\label{eq:2.4}
\end{equation}
in the case of bond-type disorders~\cite{kyhirano}.  
Here $i$ denotes the position of a disorder. 
\par
 The number of lattice points, $N$, is chosen as a multiple of four, in 
 order to have a perfectly dimerized ground state with a 
 non-degenerate top state of the valence band and with a 
 non-degenerate bottom state of the conduction band. The total number 
 of electrons is set to be equal to $N$ so that the half-filled ground 
 state may be realized. As described above, we prepare first the 
 ground state which  should  be fully self-consistent with respect to 
 both degrees of freedom of the electrons and lattice displacements. 
 The ground state is obtained by iteration as in previous 
 works~\cite{bib:I,bib:II,bib:III,bib:IV}.
 The iteration is repeated until the ratio between the two sums, 
$\displaystyle\sum_n(y_n^{(i+1)}-y_n^{(i)})^2$ and 
 $\sum_n(y_n^{(i+1)})^2$, becomes less than $10^{-15}$, where 
 $y_{n}^{(i)}$ indicates the value of $y_{n}$ at the $i$-th iteration. 
 After having obtained the perfectly dimerized ground state we change 
 the occupation numbers of the valence band top state and the 
 conduction band bottom state so as to get the initial state with an 
 electron-hole pair. The equation 
of motion for the bond variable $y_{n}(\equiv u_{n+1}-u_{n})$ is expressed 
in the form,  
\begin{equation} 
M\ddot{y}_{n} = F(t), 
\end{equation} 
where 
\begin{eqnarray}
&F(t) &= K (y_{n + 1} - 2y_{n} + y_{n - 1}) \nonumber \\  
 &+& 2\alpha {\rm Re}\left[(B_{n + 1,n + 2}
 - 2B_{n,n + 1} + B_{n - 1,n})\right], 
\end{eqnarray}
with 
\begin{equation}
 B_{n n^{\prime}}={\displaystyle\sum_{\nu,s}}^\prime \Psi^{\ast}_{\nu s}(n,t) 
\Psi_{\nu s}(n^\prime,t). 
\end{equation} 
The prime attached to the summation symbol denotes the sum over states 
occupied by electrons in the initial state. In fact the orbital 
state index $\nu$ of the time dependent electronic wave function 
$\Psi_{\nu s}$  refers to the electronic eigenstate in the 
perfectly dimerized ground state. It is clear that, in general, $\Psi_{\nu s}$'s are not 
eigenstates of the Hamiltonian at each instance. The above equation of 
motion for ${y}_{n}$ can be integrated numerically by discretizing the 
time variable with a step $\Delta t$. 
 For the numerical simulation, we suppose a periodic boundary condition, 
 i.e., $y_{N+1}=y_{1}$, $\Psi_{\nu,s}(N+1)=\Psi_{\nu,s}(1)$. 
 The periodic boundary condition avoids the end point effect.
\par
The time evolution of the electronic wave functions is calculated through 
the time-dependent Schr\"odinger equation, which can be numerically treated by 
the fractal decomposition method for exponential operators developed 
by Suzuki~\cite{suzuki,toprog,activ}, choosing the time mesh $\Delta t$ 
sufficiently small so that the variation of ${y}_{n}(t)$ during 
$\Delta t$ is very small in the electronic scale.  
\par
The values of parameters used in the present work are as follows; 
$t_{0} =$ 2.5 eV, $K =$ 21 eV/\AA$^2$, ${\alpha}=4.1$ eV/\AA, 
$a=1.22$ \AA.  These parameters correspond to a dimerization amplitude 
$\left|{y_{0}}\right| = 0.078$ \AA, the band gap 
$2\Delta_{0} (= 4\alpha |y_{0}|)=$1.30 eV
and the bare optical phonon energy $\hbar \omega_{Q} ( = \sqrt {4K/M})$ 
= 0.16 eV. The time mesh is set to be 
$\Delta t = 0.0025\ \omega_Q^{-1} (= 0.01\ {\rm fs})$ throughout the 
paper.
\par
In the present simulation the disorder expressed by eq.~(\ref{eq:2.3}) 
or eq.~(\ref{eq:2.4}) is introduced to initiate the 
soliton-anti-soliton pair creation. In order to reduce the extrinsic 
cause as small as possible, we choose the intensity of disorder to be 
quite small (see the next section), although the actual value of the 
disorders due to dopants or bond disorders might be much 
larger~\cite{bib:ohfuti,hari91,yamashiro}.
\section{Results of Simulations}
In what follows, the bond variable $y_{n}$ and the square amplitude of 
the electron wave function are express in terms of the smoothed 
expressions according to the following definitions, 
\begin{eqnarray}
\overline{y_{n}} &=&
{\frac{1}{4}}\{(-)^{n+1}y_{n+1}+2(-)^{n}y_{n}+(-)^{n-1}y_{n-1}\}, \nonumber \\  
& &
\label{variable}
\end{eqnarray}
\begin{eqnarray}
\overline{|\Psi_{\nu s}(n,t)|^{2}} =
\frac{1}{4}
&\{&
  {|\Psi_{\nu s}(n+1,t)|}^{2} 
+2{|\Psi_{\nu s}(n,t)|}^{2} \nonumber \\  
 &+& {|\Psi_{\nu s}(n-1,t)|}^{2} 
\}.  
\label{density1}
\end{eqnarray}
\subsection{Results for the site type disorder}
In Fig.~\ref{SSoliton} we show typical examples of simulations in the 
form of $\overline{y_{n}(t)}$. In all the examples a single dopant 
with $V_{\rm s}=-0.5\times 10^{-3}t_{0}$ is placed at the $60$-th site. 
The system size is $N=120$, $132$, $144$ and $156$ from (a) to (d), 
respectively. Other conditions are the same for all the cases. The 
common feature is a couple of initial uniform oscillations which 
continue up to $t \sim 30\omega_{Q}^{-1}$. The time of this initial 
uniform oscillation seems to depend on the strength of the dopant 
potential $V_{\rm s}$. Although we did not perform any quantitative analysis, the 
initial oscillation time is found to get longer for a weaker potential.
\par
After the initial uniform oscillations, the $n$-dependence of 
$\overline{y_{n}(t)}$ begins to show some structures. In (b) and (c), 
two solitons seem to be formed and to move in opposite directions, 
while in (a) and (d), a bound state of the two solitons seems to be 
created. In the former case a localized oscillating excitation which 
might be regarded as a breather is left between the two 
solitons~\cite{breather1,breather2}. It is not 
systematic but rather stochastic in a certain sense whether a soliton 
pair moving in the opposite directions is created or the created two 
solitons form a bound pair. By changing 
the system size the energy per site supplied through the electron-hole 
excitation is changed. A slight difference of this energy might have 
affected which channel of the relaxation process should be chosen.
\par
The period of the uniform oscillation of the bond variable in the 
initial stage is estimated to be $11 \omega_{Q}^{-1}$. Even after the 
formation of a soliton pair this uniform oscillation still 
remains, though its period become a little bit shorter 
($\sim 10 \omega_{Q}^{-1}$).
This period is thought to correspond to that of the renormalized 
optical mode, the frequency of which is known to be 
$\omega_{0}=0.63\omega_{Q}$ 
($2\pi/\omega_{0}\simeq 10\omega_{Q}^{-1}$)
from the linear mode analysis for the ground 
state. The reason why the period of the initial uniform oscillation is 
slightly longer than $2\pi/\omega_{0}$ is not clear. Presumably, the 
fact that the electron system is not in the lowest energy state would 
be one of possible reasons. The period of the corresponding oscillation 
after the formation of the soliton pair become very near to 
$2\pi/\omega_{0}$. This indicates that the electron system is one of the lowest 
energy states in the presence of the soliton type 
deformations. Since the soliton is a localized object, 
the presence of a soliton does not change the frequency of the 
optical mode as far as the electronic configuration is of the lowest 
energy.
\par
In order to clarify the motion of the soliton pair, we 
show the contourgraphic presentation of $\overline{y_{n}(t)}$ in 
Fig.~\ref{Contourgraphic}.
(a) to (d) correspond to those in Fig.~\ref{SSoliton}, though the 
data for longer time are depicted. From this figure the speeds of the 
solitons in (b) and (c) are estimated to be $2.75 v_{\rm s}$ and $2.48 
v_{\rm s}$, respectively, where $v_{\rm s}$ (=$\omega_{Q}a/2$) is the 
sound velocity of the system. In both cases the right-going and left-going 
solitons have different speeds. The estimated value is the average speed, 
of the right- and left-going solitons. The fact that the 
speed of the solitons in (c) is a slightly smaller than that 
in (b) would be understand if we notice the initial energy supply per 
site is smaller in (c) than in (b).
\par
In the case of the bound soliton pair formation, Fig.~\ref{Contourgraphic} (a) 
and (d), the relative distance between bound solitons oscillates with 
a frequency about $0.63\omega_{Q}$ which is near the breather 
frequency estimated in refs.~\citen{breather1} and ~\citen{breather2}, 
although the breather like oscillation 
looks not very stable. It should be noted that the lattice vibrations 
are spreading from the bound soliton pair region.
\par
Furthermore, in order to see the electronic structure in the case of 
Fig.~\ref{SSoliton} (b) (or Fig.~\ref{Contourgraphic} (b)), we have 
plotted in Fig.~\ref{leveldensity} the snapshot of the contributions 
to the electron density from the states, 
$\psi_{\rm c}$ and $\psi_{\rm v}$, which were originally at the 
bottom of the conduction band and at the top of the valence band, 
respectively. The last one in Fig.~\ref{leveldensity} is expressed as
\begin{equation}
\tilde{\rho}(n,t)={\displaystyle\sum_{\nu,s}}^{\prime \prime}
\overline{|\Psi_{\nu s}(n,t)|^{2}-1}, 
\nonumber
\end{equation}
where the double prime on the summation symbol means the sum over all 
the valence band states except for the top state, the state indices 
representing those in the initial state. The selected time is 
$t=100\omega_{Q}^{-1}$. For reference the bond variable at that time 
is also depicted. It will be easily understood that $\psi_{\rm c}$ and 
$\psi_{\rm v}$ are linear combinations of two mid-gap states located at 
the positions of two solitons, respectively. The holes seen in 
$\tilde{\rho}$ amount to one electron for each soliton, total two 
electrons. The integrations of $|\psi_{\rm c}|^{2}$ and $|\psi_{\rm v}|^{2}$ 
yield one electron, respectively. The sum of
$|\psi_{\rm c}|^{2}$, $|\psi_{\rm v}|^{2}$ and 
$\tilde{\rho}$ is almost unity everywhere. This fact does not 
necessarily mean that the solitons are neutral, as will be discussed in the 
following.
\par
In the present calculation, the spin of electrons is not treated explicitly 
except for the degeneracy. In the initial state the occupation numbers of the 
top of the valence band and the bottom of the conduction band are equal to 
unity, other valence band states being doubly occupied and other 
conduction band states being empty. The spin state is irrelevant since the 
electron-electron interactions are neglected. In this situation it will be 
plausible to assume the two spin states are realized with an equal probability. 
This means there will be no net spin density. 
Based on the symmetry argument Ball, Su and Schrieffer~\cite{bss} showed that 
the branching ratio of the neutral soliton pair formation versus the charged 
pair formation vanishes at least in the presence of the charge conjugation 
symmetry (or the electron-hole symmetry). This result is consistent with 
experiments.~\cite{flood1,flood2} Even in the presence of the electron 
mutual interactions, namely in the absence of the charge conjugation symmetry 
the aforementioned branching ratio is negligibly small as argued by Kivelson 
and Wu.~\cite{kiv-wu} Although in the present work a weak disorder potential 
violating the charge conjugation symmetry is introduced, the result obtained 
in ref.~\citen{bss} would be still valid.
\par
After all these considerations, it is most natural to regard the result shown 
in Fig.~\ref{leveldensity} as indicating that the two branches of charged 
soliton pair states are realized with equal weight. Here the two branches 
mean two different configurations of two charged solitons, a positive one on 
the left side and a negative one on the right side or vice versa. Kivelson 
and Wu discussed the scenario that the charged soliton pairs decay into 
neutral soliton pairs based on the fact that the formation energy of a 
neutral pair is smaller than that of a charged pair in the presence of 
electron mutual interactions. Since in the present treatment the electron 
interactions are not included, it will be plausible to assume such a 
decay process is not seen. 
\par
In order to see the structures of $\psi_{\rm c}$ and $\psi_{\rm v}$ more closely, we 
investigated the time development of ${\rm i}^{n} \psi_{\rm c}$, $(-{\rm 
i})^{n} \psi_{\rm c}$, 
${\rm i}^{n} \psi_{\rm v}$ and $(-{\rm i})^{n} \psi_{\rm v}$; the factor $(\pm {\rm i})^{n}$ is 
multiplied to take out the component oscillating with a wave number 
$\pm k_{\rm F} (=\pm \pi/2a)$. 
An example of a snapshot at $t=100\omega_{Q}^{-1}$ is shown in 
Fig.~\ref{excitedsoliton} again for the case of Fig.~\ref{SSoliton}(b) 
(or Fig.~\ref{Contourgraphic}(b)). 
The smoothing procedure is applied according to eq.~(\ref{density1}). 
If we express the mid-gap states 
related to the kinks around $n=30$ and $n=90$ by real localized wave functions 
$\phi_{\rm L}$ and $\phi_{\rm R}$, respectively, the results shown in Figs.~\ref{excitedsoliton}(a) 
and (b) indicate that, apart from normalization factors, the following 
relations are roughly satisfied,
\begin{eqnarray}
{\rm Re}\overline{[{\rm i}^{n}\:\psi_{\rm c}]}&\propto& -\overline{\phi_{\rm R}},\nonumber \\
{\rm Im}\overline{[{\rm i}^{n}\:\psi_{\rm c}]}&\propto& -\overline{\phi_{\rm L}},\nonumber \\
{\rm Re}\overline{[{\rm i}^{n}\:\psi_{\rm v}]}&\propto& \;\overline{\phi_{\rm R}},\nonumber \\
{\rm Im}\overline{[{\rm i}^{n}\:\psi_{\rm v}]}&\propto& -\overline{\phi_{\rm L}}. 
	\label{eq:3.3}
\end{eqnarray}
which means that $\psi_{\rm c}$ and $\psi_{\rm v}$ are expressed as follows within very 
crude approximation,
\begin{eqnarray}
\overline{\psi_{\rm c}}&\propto&-(-{\rm i})^{n} \overline{\phi_{\rm R}}
                  -{\rm i}\:(-{\rm i})^{n} \overline{\phi_{\rm L}},\nonumber \\
\overline{\psi_{\rm v}}&\propto& \;(-{\rm i})^{n} \overline{\phi_{\rm R}}
                   -{\rm i}\:(-{\rm i})^{n} \overline{\phi_{\rm L}}. 
	\label{eq:3.4}
\end{eqnarray}
A similar consideration based on Figs.~\ref{excitedsoliton}(c) and (d) leads to the 
following relations,
\begin{eqnarray}
\overline{\psi_{\rm c}}&\propto& \;{\rm i}^{n} \overline{\phi_{\rm R}}
                  -{\rm i}\:{\rm i}^{n} \overline{\phi_{\rm L}},\nonumber \\
\overline{\psi_{\rm v}}&\propto& -{\rm i}^{n} \overline{\phi_{\rm R}}
                   -{\rm i}\:{\rm i}^{n} \overline{\phi_{\rm L}}.
	\label{eq:3.5}
\end{eqnarray}
Apparently different expressions eqs.~(\ref{eq:3.4}) 
and~(\ref{eq:3.5}) can be 
consistently satisfied if $\psi_{\rm c}$ and $\psi_{\rm v}$ (without smoothing) are 
expressed as
\begin{eqnarray}
	\psi_{\rm c}&\propto& -{\rm i}\:{\rm sin}(\frac{n \pi}{2})\phi_{\rm R}(n)
                    -{\rm i}\:{\rm cos}(\frac{n \pi}{2}) \phi_{\rm L}(n),\nonumber \\
	\psi_{\rm v}&\propto&{\rm i}\:{\rm sin}(\frac{n \pi}{2})\phi_{\rm R}(n)
                	-{\rm i}\:{\rm cos}(\frac{n \pi}{2}) \phi_{\rm L}(n), 
	\label{eq:3.6}
\end{eqnarray}
and the mid-gap states $\phi_{\rm R}(n)$ and $\phi_{\rm L}(n)$ are non-vanishing only at 
odd $n$ and even $n$, respectively. Close inspection of numerical data shows 
that the last condition is approximately satisfied. It is a common 
understanding that the mid-gap state for a static soliton vanishes at every 
second site in the absence of electron-electron interactions. Thus it is 
plausible to assume the relations shown in eq.~(\ref{eq:3.6}) to be satisfied.
\par
The factors ${\rm sin}(n\pi/2)$ and ${\rm cos}(n\pi/2)$ in 
eq.~(\ref{eq:3.6}) indicate that 
both of $\psi_{\rm c}$ and $\psi_{\rm v}$ involve $k_{\rm F}$ and 
$-k_{\rm F}$ 
components with an equal weight. $\psi_{\rm v}$ and $\psi_{\rm c}$ can be regarded also 
as bonding and anti-bonding combinations of two mid-gap states $\phi_{\rm R}$ and 
$\phi_{\rm L}$. This fact may be most easily seen by calculating a product 
$\psi_{\rm c} \psi_{\rm v}$ which is effectively expressed as
\begin{equation}
\psi_{\rm c}\psi_{\rm v}\propto[\phi_{\rm R}(n)]^{2}-[\phi_{\rm L}(n)]^{2}
	\label{eq:3.7}
\end{equation}
In fact we have checked this is satisfied at least for the data shown in 
Fig.~\ref{excitedsoliton}. 
The above results are consistent with our notion that the 
two different combinations of a pair of charged solitons  
are realized with an equal probability. This is 
confirmed also from the electron density profile as discussed already 
in connection to Fig.~\ref{leveldensity}. The fact that $\psi_{\rm v}$ 
and $\psi_{\rm c}$ correspond to bonding and anti-bonding linear 
combinations of $\phi_{\rm R}$ and $\phi_{\rm L}$ can be seen from the 
behavior of the energy. The one-particle energy corresponding to 
$\psi_{\rm v}$ is slightly negative, while that corresponding to 
$\psi_{\rm c}$ is slightly positive, thought both of them are very 
near the gap center. 
\par
In order to see the behavior of the system at the initial stage of the soliton 
pair formation, we show in Fig.~\ref{site-density} snap shots 
(at $t=0, 20 \omega_{Q}^{-1}$ and $40 
\omega_{Q}^{-1}$) of the bond variable $\overline{y_{n}}$ and the excess electron 
densities $\tilde{\rho_{\uparrow}}$ and $\tilde{\rho_{\downarrow}}$ 
for up and down spins defined by 
\begin{eqnarray}
\tilde{\rho}_{\uparrow}&=&{\displaystyle\sum_{\nu}}^{\prime \prime}
\overline{|\Psi_{\nu \uparrow}(n,t)|^{2}}-\frac{1}{2},\nonumber \\
\tilde{\rho}_{\downarrow}&=&{\displaystyle\sum_{\nu}}^{\prime \prime \prime}
\overline{|\Psi_{\nu \downarrow}(n,t)|^{2}}-\frac{1}{2},
\end{eqnarray}
where $\Sigma_{\nu}^{''}\cdots$ means the sum over occupied up spin states and 
$\Sigma_{\nu}^{'''}\cdots$ the sum over occupied down spin states. 
So far we did not specify the spins of the excited electron and hole. 
When drawing Fig.~\ref{site-density}, 
we have chosen the initial occupation numbers so that there is a 
conduction electron 
and a hole in the valence band for up spin, and all the valence 
band states are occupied and the conduction band is empty for down spin. The 
electron densities have small structure even at $t=0$ due to the presence of 
the disorder potential. This small structure acts as a nucleation seed for 
the soliton pair formation. It should be noted that there is essentially no 
initial structure in the bond variable.
\par
\subsection{Results for the case with a  bond type disorder}
Here we show the results in the presence of a bond type disorder,
 which might mimic the mixture of a $cis$-segment in {\it trans}-polyacetylene. 
There are four different ways of putting a bond type disorder in the 
system.
These ways correspond to four different combinations of whether the disorder 
is put at a single (long) bond or at a double (short) bond and whether the 
sign of the disorder is positive or negative.
\par
In Fig.~\ref{Bond-order}, typical results for the four different cases are depicted; the 
system size is fixed at $N=132$. The main difference from the results for the 
case with the site type disorder is that the nucleation position does not 
necessarily coincide with the disorder position but depends on the location and 
the sign of the bond type disorder. This behavior can be understood by 
considering whether the bond disorder is  ``match-type'' or  ``mismatch-type'' in 
the initial dimerized state~\cite{kyhirano}. When the disorder is match-type, 
the nucleation starts at the most distant bond from the disordered one. 
One the other hand when the disorder is mismatch-type, the nucleation 
begins at the disordered bond. This will be reasonable since    a    bond with 
a match-type disorder is stable against reversing the bond variable while a 
bond with a mismatch-type disorder is unstable. Furthermore the free motion of 
solitons is not seen. This is also reasonable if we remember the bond disorder 
yields an effective step like potential for a soliton as discussed in 
ref.~\citen{kyhirano}. In the present situation, solitons tend to move towards the 
direction where the effective potential due to the disorder increase, and this 
is thought to be the reason why solitons stop at a certain distance. 
Since the bond disorder gives a direct effect on the bond variable, the time 
before the formation of a soliton pair is shorter than 
that in the case of a site disorder with the same disorder strength.
\par
The detailed time variation of the bond variable at the early stage is shown in 
Fig.~\ref{bond-order2d} where $\Delta y_{n}=y_{n}-y_{1}$ is used 
instead of $y_{n}$ in order to remove the uniformly oscillating part. 
When the initial value of $\overline{\Delta y_{n}}$ is enhanced at the disordered 
bond, it means that the bond disorder is match-type. On the other hand, when 
it is reduced, the disorder is mismatch-type. It will be clear that the 
soliton pair formation is triggered at the bond where the initial 
value of $\overline{\Delta y_{n}}$ is smallest. When the 
disorder is mismatch-type, the inversion of the bond variable begins at the 
disordered bond, while, when the disorder is match-type, it starts at the most 
distant bond from the disordered one.   
\section{Concluding Remarks}
Starting from an electron and a hole in the uniformly dimerized background, 
the lattice relaxation process related to the soliton-pair formation in 
polyacetylene is numerically studied using the SSH model. Two kinds of 
local small disorders are introduced in order to trigger the soliton-pair 
formation; a site disorder and a bond disorder. In the case of the site 
disorder, the distortion of the bond variable which is necessary to create 
a soliton-pair begins at the disordered site irrespectively of the sign of 
the disorder potential. Initially the excess electronic energy is transferred 
to the lattice system through the electron-lattice coupling and a uniform 
oscillation of the bond variable continues up to $t \sim 30\omega_Q^{-1}$ 
in the case with the disorder potential strength 
$|V_{\rm s}|=0.5\times 10^{-3}t_0$. The period of this oscillation is 
nearly equal to that of the renormalized optical mode oscillation. The time 
length of the initial uniform oscillation of the bond variable increases 
with decreasing the potential strength $|V_{\rm s}|$, tending to be infinite 
in the limit $|V_{\rm s}| \rightarrow 0$. 
This initial uniform oscillation was not seen in the simulations in 
ref.~\citen{breather2}, which is considered to be due to the difference in 
the treatment of electronic dynamics. 
After this uniform oscillation 
the bond variable begins to deform at the disordered site. 
Although the 
deformation of the bond variable suggests that 
 two solitons have been formed, we find no local charge or spin. 
 This is thought to be due to the fact that 
two different channels of a charged soliton pair formation are realized with an 
equal weight. 
This was confirmed in terms of the behaviors of mid-gap states. 
Although it is not clear only from the results of present 
simulations whether the channels for the neutral soliton pair formation is 
suppressed or not, it will be reasonable to assume the suppression of those 
channels based on the symmetry argument by Ball, Su and Schrieffer.~\cite{bss}
It looks rather stochastic whether the created two solitons move almost freely 
to the opposite directions or they form a bound pair. The bound pair state 
seems not very stable and is dissociated into two free solitons after some time, 
though it is again stochastic when it is dissociated.
\par
In the case of a bond type disorder, the starting position of the bond order 
distortion changes depending on whether the bond disorder is located at the 
double (short) bond or at the single (long) bond and whether the sign of the 
disorder is positive or negative. This behavior was found to be explained 
by the concept of ``match'' and ``mismatch'' characteristics of the bond 
disorder. Since the bond disorder affects the bond variable directly, the 
time length of initial uniform oscillation is generally shorter for the case 
of a bond type disorder compared to the case with a site type disorder as far 
as the disorder strengths are the same order of magnitude. 
Within the knowledge of the present authors, there has been so far no precise 
analysis of the effect of bond disorders on the soliton pair formation. 
\par
In the present work, the disorder strength has been chosen to be very small 
in order to study intrinsic properties as far as possible. If we introduce 
a  strong site disorder, 
then the trapping of one of the solitons at the disorder 
site can occur. Furthermore the electron-electron interactions has been 
neglected for the sake of simplicity. If we take into account the electron 
correlation, the degeneracy between the neutral and charged solitons is 
lifted. It will be interesting to investigate the effect of this nondegeneracy 
on the soliton pair formation. This will be studied in a future work. 
\par
In the real photoexcitation of a soliton pair, the non-uniformity might be 
created during the excitation of an electron-hole pair, e.g. due to a 
non-uniformity of the light intensity. In fact, if we choose a wave-packet 
type excitation of the electron-hole pair instead of the plane wave type 
excitation as treated in the present paper, the soliton pair formation 
becomes possible even without any disorder.~\cite{shimizu} Photoexcitation 
experiments using a light source intentionally made non-uniform may be 
interesting in this sense. 
\section*{Acknowledgments}
The authors thank Dr. M. Kuwabara for useful comments. 
This work was partially financed by a Grant-in-Aid for Scientific 
Research from the Ministry of Education, Science and Culture, No. 05640446. 
A part of numerical calculations were performed on Facom VPP500 of 
Institute for Solid State Physics, University of Tokyo. 
   
\def\jpsj{J. Phys. Soc. Jpn. }
\def\ptp{Prog. Theor. Phys. }
\def\prl{Phys. Rev. Lett. }
\def\pr{Phys. Rev. }

\newpage
\begin{fullfigure}
	\caption{Stereographic presentation of the site and time dependence of 
	the bond variable(smoothed as in eq.~(\ref{variable})); the  
	system size $N$ is (a)120, (b)132, (c)144 and (d)156. 
    A single dopant with $V_{\rm s}=-0.5\times 10^{-3}t_{0}$ 
	is set at the 60-th site for all the cases.  
        The discretized time mesh $\Delta t$ is $0.0025\omega_{Q}^{-1}$.
    In (b) and (c) two solitons are excited. The excitations seen in 
    (a) and (d) look like a bound pair of two solitons.}
	\label{SSoliton}
\end{fullfigure}
\begin{figure}
	\caption{Contourgraphic presentation of the site and time dependence of 
	the bond variable(smoothed as in eq.~(\ref{variable})); the  
	system size $N$ is (a)120, (b)132, (c)144 and (d)156. 
    A single dopant with $V_{\rm s}=-0.5\times 10^{-3}t_{0}$ 
	is set at the 60-th site for all the cases. 
        The discretized time mesh $\Delta t$ is $0.0025\omega_{Q}^{-1}$.
    In (b) and (c) two freely moving solitons are excited. The excitations seen in 
    (a) and (d) are bound soliton pairs showing breather like 
    oscillations. 
    Light shade is upper side and dark shade is 
	downward side.}
	\label{Contourgraphic}
\end{figure}
\begin{figure}
	\caption{Snapshot of contributions to the excess 
electron density from different states for a 
system with the total site number $N$=132 and the up and down spin electron numbers 
$N_{\rm e\uparrow}=66$ and $N_{\rm e\downarrow}=66$ satisfying periodic boundary condition. 
(a) is the bond variable. (b) and (c) are the contributions from the 
states initially at the bottom of the conduction band and at the top 
of valence band, $\psi_{\rm c}$ and $\psi_{\rm v}$, respectively. (d) 
is the density of total electrons except for those two electrons.  
The discretized time mesh $\Delta t$ is $0.0025\omega_{Q}^{-1}$. 
A disorder potential is set at the $60$-th site and its intensity is 
$V_{\rm s}=-0.5\times 10^{-3}t_{0}$. 
The time is $t=100\omega_{Q}^{-1}$.}
	\label{leveldensity}
\end{figure}
\begin{figure}
	\caption{Snapshot of the smoothed wave functions. 
The system size $N$=132 and the time is $t=100\omega_{Q}^{-1}$.
(a) is ${\rm i}^{n} \psi_{\rm c}$, 
(b) ${\rm i}^{n} \psi_{\rm v}$, 
(c) $(-{\rm i})^{n} \psi_{\rm c}$, 
 and (d) $(-{\rm i})^{n} \psi_{\rm v}$; the factor $(\pm {\rm i})^{n}$ is 
multiplied to take out the component oscillating with a wave number 
$\pm k_{\rm F} (=\pm \pi/2a)$. 
The smoothing procedure is for example, $\overline{{\rm i}^{n} 
\psi_{\rm c}}=
{\frac{1}{4}}\{{\rm i}^{n-1} \psi_{\rm c}(n-1)+2\;{\rm i}^{n} 
\psi_{\rm c}(n)+{\rm i}^{n+1}
\psi_{\rm c}(n+1)\}$ in the case of (a). 
Solid and broken lines represent real and imaginary parts, 
respectively.  
Intensity of the disorder potential located at the 60-th 
site is $V_{\rm s}=-0.5\times 10^{-3}t_{0}$. }
	\label{excitedsoliton}
\end{figure}
\begin{fullfigure}
	\caption{
The time variation of the bond variable (the broken line, arbitrary scale) and the excess 
electron densities for up (the thick line) and down (the thin line) spins, 
$\tilde{\rho}_{\uparrow}$ and $\tilde{\rho}_{\downarrow}$, 
  in the presence of a site disorder at the 60-th site of a system with a size 
$N=132$. The figures on the left and right hand sides show the cases with 
$V_{\rm s}=0.5\times 10^{-3} t_{0}$ and $-0.5\times 10^{-3} t_{0}$, 
respectively. 
Since the excess electron densities are quite small for small 
$|V_{\rm s}|$, an enhancement factor $t_{0}/|V_{\rm s}|$ is 
multiplied to them.}
	\label{site-density}
\end{fullfigure}\par
\begin{fullfigure}
	\caption{Stereographic presentation of the $n$ and $time$ dependence of 
	the smoothed bond variable 	(eq.~(\ref{variable})).  
	The system size is $N=132$. The bond disorder is located at the 60-st 
bond (=double bond) in the case of (a) and (b) as depicted in (i), and at the 
61-th bond (=single bond) in the case of (c) and (d) as depicted in (ii). The 
bond disorder factor $V_{\rm b}$ is $-0.5\times 10^{-3} t_{0}$ for (a) 
and (c), 
and $0.5\times 10^{-3} t_{0}$ for (b) and (d). The formation of a soliton pair 
begins around $t= 10\omega_{Q}^{-1}$.} 
	\label{Bond-order}
\end{fullfigure}
\begin{fullfigure}
	\caption{
The time development of the bond variable at the early stage. The system size 
is $N=132$. Figures on the left and right hand sides show the cases with a 
bond disorder at the 84-th (double) and 83-rd (single) bonds, respectively. 
Thick lines indicate the cases with $V_{\rm b}=-0.5\times 10^{-3} t_{0}$ and thin 
lines the case with $V_{\rm b}=0.5\times 10^{-3} t_{0}$. The broken line shows the 
case without disorder for comparison. 
In order to remove the uniform oscillation we use $\overline{\Delta y_{n}}=y_{n}-y_{1}$
 instead of $y_{n}$. 
Since the variation of 
$\overline{\Delta y_{n}}$ is quite 
small when $|V_{\rm b}|$ is small, an enhancement factor 
$t_{0}/|V_{\rm b}|$ is 
multiplied to $\overline{\Delta y_{n}}/a$.} 
	\label{bond-order2d}
\end{fullfigure}
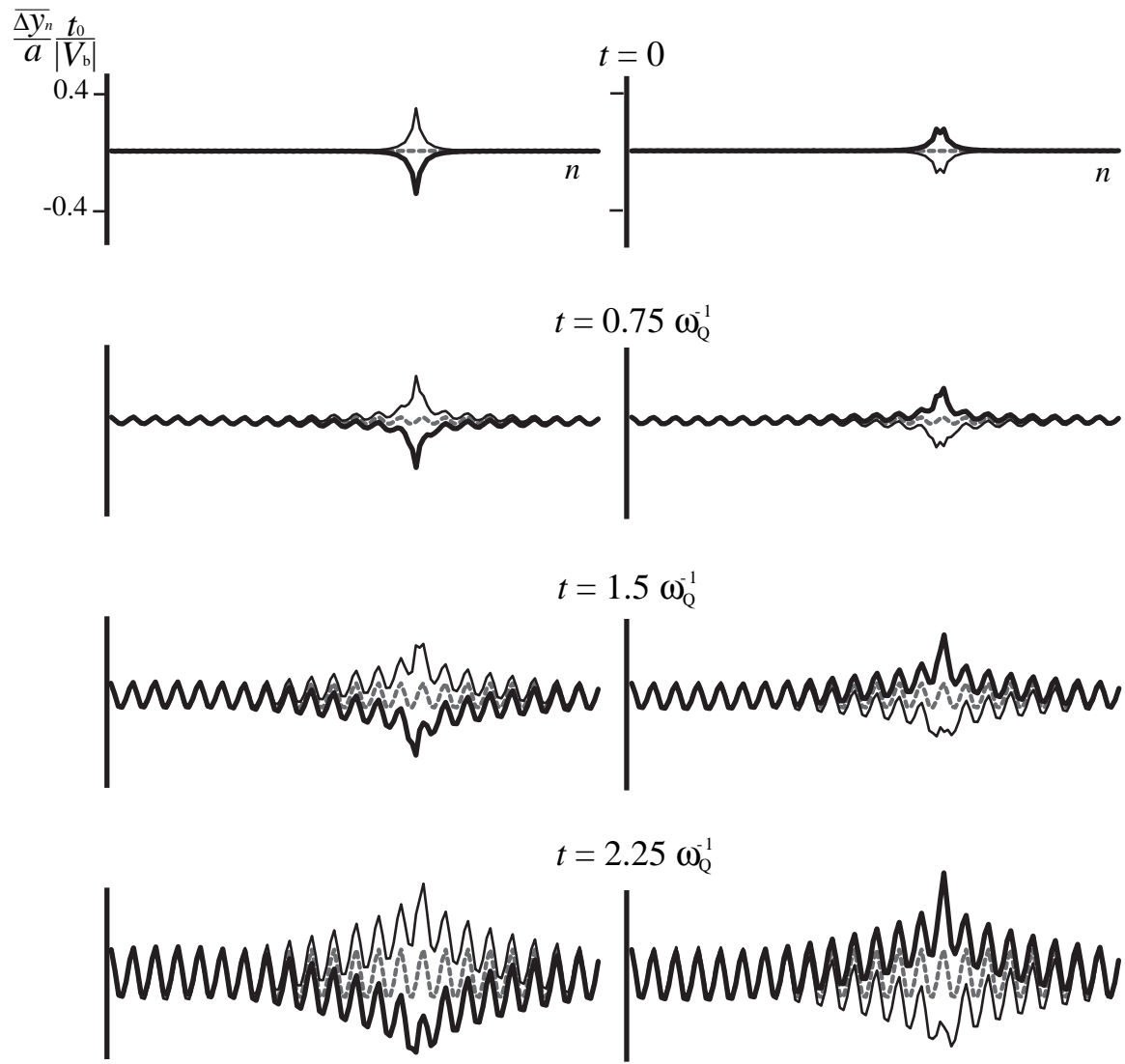
\end{document}